\documentclass[fleqn,twoside,twocolumn,nofootinbib,showkeys]{revtex4} 
\usepackage[sec,nocpr]{ujp} 

\begin{document}
\title[Elastic strains in SiGe heterostructures]
{ELASTIC STRAINS IN SiGe HETEROSTRUCTURES\\ WITH NON-UNIFORM QUANTUM
DOTS}%
\author{V.V. Kuryliuk}
\affiliation{Taras Shevchenko National University of Kyiv, Faculty of Physics}
\address{64/13, Volodymyrs'ka Str., Kyiv 01601, Ukraine}
\email{kuryluk@univ.kiev.ua}

\udk{???}  \razd{\secviii}

\autorcol{V.V.\hspace*{0.7mm}Kuryliuk}

\setcounter{page}{780}%

\begin{abstract}
Elastic strain distributions in SiGe heterostructures with quantum
dots have been simulated with the use of the finite element method. The effect
of a non-uniform germanium distribution in the nanoislands on the
spatial dependence and the magnitude of elastic fields was studied. It
is shown that quantum dots with a uniform component content are
more strained in comparison with non-uniform nanoislands.
\end{abstract}
\maketitle

\section{Introduction}\vspace*{2mm}

The unique electronic and optical properties of silicon--germanium
heterostructures with self-organized quantum dots (nanoislands) create a basis
for their practical application as promising materials to\ modern nano- and
optoelectronics \cite{1}. In particular, a high absorption factor of those
heterostructures in the range of energies lower than the germanium energy gap
width is successfully used in the manufacture of infra-red radiation detectors
of a new generation \cite{2}. Moreover, silicon--germanium structures can serve
as active elements in solar batteries \cite{3} and light emitting \cite{4} and
spintronic \cite{5} devices.

The properties of silicon heterostructures Si$_{1-x}%
$Ge$_{x}$ with quantum dots, where $x$ is the germanium content in the
compound, obtained following the Stranski--Krastanov growth mode are
closely connected with elastic deformations and accompanying
mechanical stress fields in the structures. It is the elastic fields
arising owing to the mismatch between the material lattices that
play a crucial role in the growing of a heterostructure, being
responsible for a spatial ordering of nanoislands and their shape
\cite{6}. Moreover, elastic strains substantially affect the band
structure (the confinement-potential for charge carriers) of
crystals, the mobility and the effective masses of electrons and
holes in them, and, hence, change the properties of heterostructures
on the whole \cite{7}. Therefore, an important task of the physics and
the technology of nanodimensional semiconducting structures consists in
developing the methods for the determination of and the control over elastic
strain fields, as well as their manipulation by means of varying the
physical parameters of heterostructures.

Experimental researches of mechanical stresses in low-dimensional
heterostructures are based on the Raman scattering technique
\cite{8,9}. In particular, this method enables the peculiarities in
a crystal structure of strained germanium nanoislands in the silicon
matrix \cite{10} or their morphology \cite{11} to be determined.
However, the analysis of Raman spectra allows only the averaged
values of strains to be estimated and provides no information
concerning their distribution in the islands and near to them.
Taking all the above into account, the methods of computer
simulation turn out an effective tool to research nanosystems, in
particular, silicon--germanium \mbox{heterostructures.}\looseness=1

For today, there are plenty of works devoted to the study of elastic strain
fields and their influence on the properties of Si$_{1-x}$Ge$_{x}$
heterostructures with quantum dots \cite{12,13,14}. There are a number of
approaches to calculate the elastically deformed state of structures with
nanoislands; they use the molecular dynamics, Green's function, or finite
element methods \cite{15,16,17}. The shortcoming of the former consists in a
considerable computation time even if objects with small volume are
calculated. The Green's function method is used mostly if the
structures are analyzed in the approximation of infinite or semiinfinite
substrate; however, it cannot be always applied to confined systems. The most
widespread is the method of finite elements, which allows the calculations to
be carried out for objects with complicated geometry and does not require
considerable computational resources.

In the majority of the known works, where the heterostructure
properties induced by elastic fields are studied, the obtained
results are analyzed making the assumption that the component
content in nanoislands is uniform. However, in the course of
high-temperature epitaxy of germanium onto a silicon substrate, the
components become partially mixed, so that their contents change
over the quantum dot volume. As was shown in work \cite{18}, the
redistribution of the silicon and germanium concentrations in
nanoislands results in a decrease of the energy of the system, and
this redistribution therefore turns out to be energetically
beneficial. The corresponding distribution of germanium (silicon) is
related to the shape of quantum dots and the content ratio between
Si and Ge in them. Bearing in mind that the concentration gradient
can be rather substantial, the account of a non-uniformity in the
contents of island components is an important factor, while studying
the properties that are governed by the electron structure of
quantum dots, because this structure is sensitive to deformations in
\mbox{heterostructures.}\looseness=1

In this work, we used the finite element method to calculate the strain fields
in silicon heterostructures with germanium quantum dots. The calculations were
carried out in the framework of the elastic continuum model. A comparative
analysis of the results obtained for nanoislands with uniform and non-uniform
germanium distributions over their volume was made. We also analyzed the
influence of a component content non-uniformity on the band structure in
Si$_{1-x}$Ge$_{x}$ quantum dots.

\section{Procedure for Calculating Strains in~SiGe~Heterostructures}

\begin{figure}
\vskip1mm
\includegraphics[width=\column]{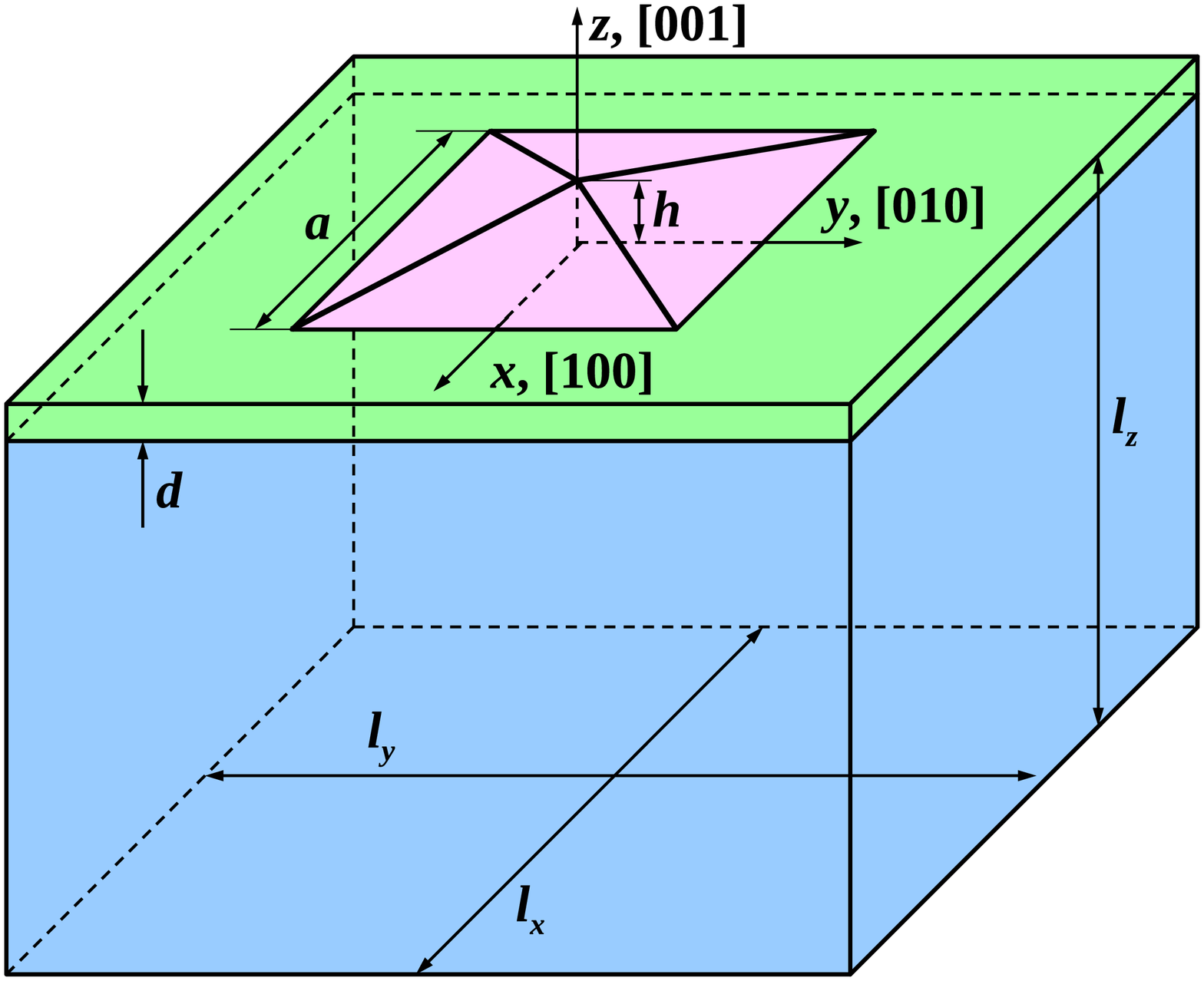}
\vskip-3mm\caption{Geometrical parameters of the studied SiGe
heterostructure }
\end{figure}

We will consider self-organized germanium quantum dots on a silicon surface as
a regular array of islands with identical dimensions. The typical period (the
distance between neighbor nanoislands) in such heterostructures equals a few
tens of nanometers. Therefore, while calculating elastic strains in the
silicon substrate--germanium quantum dot system, let us confine the consideration to
a single model cell that includes one quantum dot in the form of a regular
square pyramid centered on a substrate with the transverse dimensions
$l_{x}\times l_{y}$ and the thickness $l_{z}$ (Fig.~1). Between the Si
substrate and the germanium quantum dot, there is a wetting layer of Ge of the
thickness $d$. The height and the base side length of the pyramid equal $h$ and
$a$, respectively. The origin of a coordinate system is fixed at the center of
the pyramid base, and the $x$, $y$, and $z$ axes are directed along the
crystallographic directions [100], [010], and [001], respectively.

The strain fields in heterostructures with quantum dots will be studied in the
elastic continuum approximation with the use of the standard equations of elasticity theory,
\begin{equation}
\dfrac{\partial\sigma_{ij}}{\partial x_{j}}=0,
\end{equation}\vspace*{-7mm}%
\begin{equation}
\sigma_{ij}=C_{ijkl}\left[
\varepsilon_{kl}-\varepsilon_{0kl}\right]\!,
\end{equation}\vspace*{-7mm}%
\begin{equation}
\varepsilon_{kl}=\dfrac{1}{2}\left(\!  \dfrac{\partial u_{k}}{\partial x_{l}%
}+\dfrac{\partial u_{l}}{\partial x_{k}}\!\right)\!,
\end{equation}
where $\sigma_{ij}$, $\varepsilon_{kl}$ and $C_{ijkl}$ are the tensors of
mechanical stresses, elastic strains, and elastic moduli, respectively; and
$u_{k}$ is the elastic displacement vector. The quantities $\varepsilon_{0kl}$
in Eq.~(2) stand for initial strains originated from a mismatch between the
lattice parameters in the substrate and nanoisland materials (misfit
strains),
\begin{equation}
\varepsilon_{0ij}=\dfrac{a_{s}-a_{i}}{a_{s}}\delta_{ij},
\end{equation}
where $a_{s}$ and $a_{i}$ are the lattice constants for the substrate and the
quantum dot, respectively, and $\delta_{ij}$ is the Kronecker symbol. Since
$a_{s}=5.430~\mathrm{\mathring{A}}$ for silicon and $5.646~\mathrm{\mathring
{A}}$ for germanium, the absolute value of $\varepsilon_{0ij}$ amounts to
approximately 4\%. In the calculations, the initial strain was considered to
be nonzero only in the quantum dot. The subsequent structure relaxation
resulted in the appearance of strains in the whole substrate as well.

Notice that the lattice mismatch is not the unique source of
mechanical stresses. Since the heterostructures are grown up at
epitaxial temperatures, their subsequent cooling down may invoke
stresses stemming from the difference $\Delta\alpha$ between the
thermal expansion coefficients. The corresponding strains
$\varepsilon_{\rm TE}$ arisen in the structure can be estimated from
the relation
\begin{equation}
\varepsilon_{\rm TE}=\Delta\alpha\Delta T,
\end{equation}
where $\Delta T$ is the temperature change. We are interested in the
strains that take place in heterostructures at room temperature,
i.e. $\Delta T\simeq500~^{\circ}\mathrm{C}$. Then, taking into
account that $\Delta
\alpha=3.3\times10^{-6}~^{\circ}\mathrm{C}^{-1}$ for the heteropair
Si--Ge, we obtain the value $\varepsilon_{\rm TE}\approx0.2\%$.
Hence, the value of $\varepsilon_{\rm TE}$ is an order of magnitude
less than the strains $\varepsilon_{0ij}$, and we will consider
below only mechanical stress fields related to the lattice
mismatch between the substrate and nanoisland materials.

Equations (1)--(3) have to be appended by boundary conditions for the unknown
components of the displacement vector. In particular, in view of the
periodic character of heterostructure, the normal components of the vector $u_{k}$
were fixed at the opposite edges $x=\pm l_{x}/2$ and $y=\pm l_{y}/2$ of the model
cell,%
\begin{equation}
u_{x}\left(\! x=\pm \dfrac{l_{x}}{2}\!\right) = u_{y}\left(\! y=\pm
\dfrac{l_{y}}{2}\!\right)=0.
\end{equation}
The thickness of the wetting layer near the quantum dot, where the
strain fields are considered, is much narrower than the substrate
thickness. Therefore, the following \textquotedblleft
rigid\textquotedblright\ boundary conditions are selected at the
lower substrate edge $z=-l_{z}$:
\begin{equation}
u_{i}\left(  z=-l_{z}\right)  =0,\quad i=x,y,z.
\end{equation}
In addition, the absence of mechanical stresses is supposed at all external
surfaces, including the open surfaces of germanium nanoislands and the silicon substrate.

The formulated problem has no analytical solution. Therefore, in this work,
the elastic fields in silicon--germanium heterostructures are determined within
the finite element method. The calculations were carriued out according to
a Fortran program written by the author, in which the procedures of the Intel MKL
mathematical library were used. The model cell containing a quantum dot on a
silicon substrate was divided into a mesh of tetrahedral elements, each of
them containing 10~nodes. Within every element, the unknown components of
the elastic displacement vector were approximated by a linear combination of
the so-called shape functions $\xi\left(  x,y,z\right)  $ \cite{16},
\begin{equation}
u_{k}\left(  x,y,z\right)  =\sum_{i=1}^{10}\nu_{ik}\xi_{i}\left(
x,y,z\right)\!,
\end{equation}
where $\nu_{ik}$ are the unknown coefficients equal to $u_{k}$-values at the mesh
nodes. The application of the Galerkin method \cite{19} allowed the differential
equations (1)--(3) with the corresponding boundary conditions to be
transformed into a system of algebraic equations for the unknown coefficients
$\nu_{ik}$,%
\begin{equation}
\left[  K\right]  \left[  \nu\right]  =\left[  f\right]\!,
\end{equation}
where the components of the matrices $\left[  K\right]  $ and $\left[  f\right]  $
are determined by the relations%
\begin{equation}
K=\int\limits_{Ve}{B^{T}CBdV},
\end{equation}\vspace*{-5mm}%
\begin{equation}
f=\int\limits_{Ve}{B^{T}C\varepsilon_{0}dV}.
\end{equation}
Integration in expressions (10) and (11) is carried out over the element
volume $V_{e}$, the upper index $T$ means the transposition operation, and
$B^{T}$ stands for the operator of the following form:%
\begin{equation}
B^{T}=\begin{bmatrix} \dfrac{\partial \xi }{\partial
x}&0&0&0&\dfrac{\partial \xi}{\partial z}&\dfrac{\partial
\xi}{\partial
y}\\[3mm]
0&\dfrac{\partial \xi}{\partial y}&0&\dfrac{\partial \xi}{\partial
z}&0&\dfrac{\partial \xi}{\partial
x}\\[3mm]
0&0&\dfrac{\partial \xi}{\partial z}&\dfrac{\partial \xi}{\partial y}&\dfrac{\partial \xi}{\partial x}&0\\
\end{bmatrix}\!\!.
\end{equation}

The coefficients $\nu_{ik}$ obtained from the solution of Eqs.~(9) were used
to calculate the components of the elastic displacement vector $u_{k}$. Then,
with the use of relation (3), the strain tensor components were calculated.

\section{Results and Their Discussion}

Elastic strains in heterostructures were calculated for the
following parameters of nanoislands: the base side length
$a=30$\textrm{~nm}, height $h=4.5$\textrm{~nm}, and wetting layer
thickness $d=0.5$\textrm{~nm}. The dimensions of the model cell were
chosen to equal the average distance between quantum dots in real
structures, i.e. $l_{x}=$ $=l_{y}=60$\textrm{~nm}. To exclude the
influence of the lower substrate face, the dimension $l_{z}$ was
selected to be 10 times larger than the nanoisland height,
$l_{z}=10h=$ $=45$\textrm{~nm}.

In the calculations connected with the non-uniform content of
germanium in quantum dots, the distribution of the Ge concentration,
whose profile is depicted in Fig.~2, was used. The exhibited
dependence is similar to that obtained earlier in work \cite{18} for
pyramidal nanoislands. The maximum germanium content,
\mbox{$x\approx1$}, was observed near the island vertices, and its
minimum, $x\approx0.3$, in vicinities of the vertices near the base.
The value of germanium content averaged over the whole quantum dot
volume was about 0.7. Therefore, the data obtained for
Si$_{1-x}$Ge$_{x}$ heterostructures with non-uniform nanoislands
were compared with the results of calculations for islands with a
uniform component content, Si$_{0.3}$Ge$_{0.7}$. The values of
components of the tensor of elastic moduli for pure silicon,
$C_{ij}(\mathrm{Si})$, and germanium, $C_{ij}(\mathrm{Ge})$, were
taken from work \cite{20}. For the compound
Si$_{1-x}$Ge$_{x}$, the following linear approximation was used:%
\begin{equation}
C_{ij}(\mathrm{Si}_{1-x}\mathrm{Ge}_{x})=\left[  C_{ij}(\mathrm{Ge}%
)-C_{ij}(\mathrm{Si})\right]  x+C_{ij}(\mathrm{Si}).
\end{equation}

\begin{figure}
\vskip-3mm
\includegraphics[width=\column]{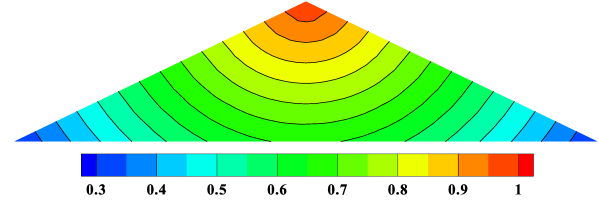}
\vskip-3mm\caption{Non-uniform distribution of the germanium content
$x$ in a pyramidal quantum dot Si$_{1-x}$Ge$_{x}$ used for the
calculations of elastic fields }\vskip3mm
\end{figure}

\begin{figure}
\includegraphics[width=\column]{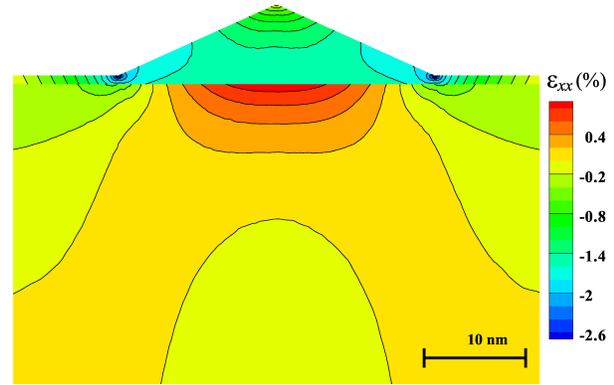}
\vskip-2mm\caption{Spatial distribution of the strain tensor
component $\varepsilon_{xx}$ in the $xz$-plane for a heterostructure
with non-uniform Si$_{1-x}$Ge$_{x}$ nanoislands  }\vskip3mm
\end{figure}

In Fig.~3, the calculated distribution of the $\varepsilon_{xx}$
component of the strain tensor in the heterostructure with a
non-uniform nanoisland content is depicted. The $\varepsilon_{xx}$
magnitude substantially varies in the quantum dot and near its base,
so that high strain gradients arise in those heterostructure
regions. One can see that the stretching strains prevail in the
silicon substrate along the Si--Si$_{1-x}$Ge$_{x}$ heterojunction,
with the maximum value $\varepsilon_{xx}=0.6\%$ attaining near the
island base. At the same time, the $\varepsilon_{xx}$ component
changes the sign in the quantum dot, and this section of the
heterostructure undergoes compressive strains along the
heterojunction with a maximum value of 2.5\% near the base angles.
The magnitude of compressive strains gradually decreases from the
pyramid base to its vertex. The component $\varepsilon_{yy}$ of the
strain tensor has a similar dependence, and its distribution is
\mbox{not shown.}

\begin{figure}
\vskip1mm
\includegraphics[width=\column]{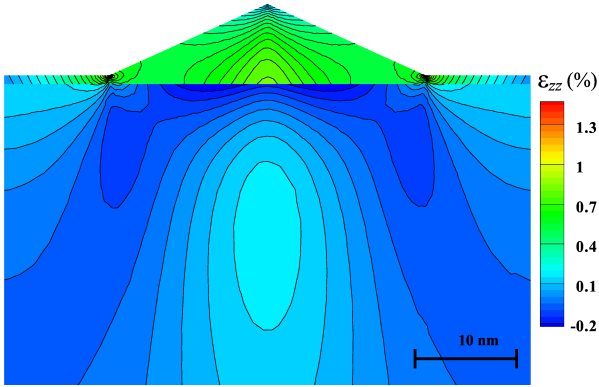}
\vskip-3mm\caption{The same as in Fig.~3, but for the strain tensor
component $\varepsilon_{zz}$  }\vskip1mm
\end{figure}
\begin{figure}
\includegraphics[width=7cm]{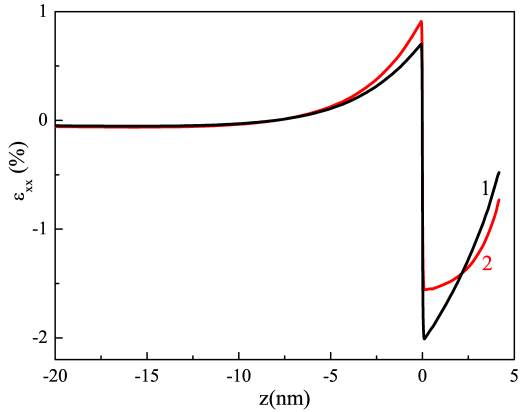}
\vskip-3mm\caption{Distributions of the strain tensor component
$\varepsilon_{xx}$ along the axis $Oz$ in heterostructures with
uniform (\textit{1}) and non-uniform (\textit{2}) nanoislands
}\vskip1mm
\end{figure}
\begin{figure}
\includegraphics[width=7cm]{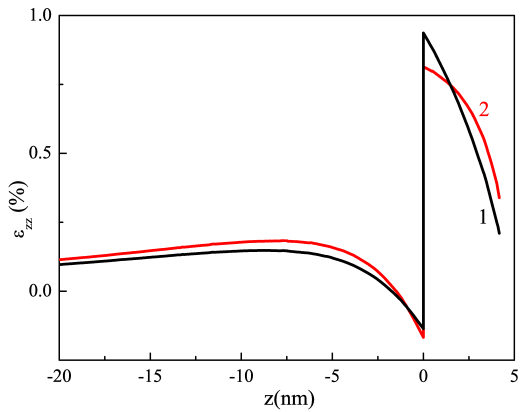}
\vskip-3mm\caption{The same as in Fig.~5, but for the strain tensor
component $\varepsilon_{zz}$}
\end{figure}

In the elastic deformation interval, the ratio between the longitudinal and
transverse elongations is constant (the Poisson effect). Therefore, the strain
component $\varepsilon_{zz}$ has the opposite sign (Fig. 4): the substrate
near the heterojunction undergoes compressive strains in the
heterostructure growth direction, whereas nanoislands undergo the action of stretching
strains. Moreover, the $\varepsilon_{zz}$ component changes its sign in the
substrate depth. At a distance of the order of the island height, the stretching
strains have the maximum value, $\varepsilon_{zz}\approx0.17\%$; then, they
gradually decrease toward the substrate depth.

Qualitatively, the distributions of $\varepsilon_{xx}$ ($\varepsilon_{yy}$)
and $\varepsilon_{zz}$ in heterostructures with islands characterized by a
uniforn component content have the same features as in the case considered
above. To elucidate the quantitative effect of non-uniformity in the
distribution of elastic stress fields over heterostructures, let us consider
the profiles of the tensor components $\varepsilon_{ii}$ along the $Oz$-axis
(Figs.~5 and 6).

As is seen from Fig.~5, the compressive strains $\varepsilon_{xx}$ in the
lower part ($z<h/2$) of non-uniform quantum dots are smaller by absolute value
than the corresponding values in uniform islands. The maximum difference is
observed near the pyramid base, being equal to $\Delta\varepsilon_{xx}%
\approx0.45\%$. Approximately at the middle of the quantum dot height, the
component $\varepsilon_{xx}$ is identical in the islands of both types and
equals $\varepsilon_{xx}\approx1.4\%$. In the upper part of a quantum dot
($z>h/2$), the nanoislands with a non-uniform germanium distribution turn out
more deformed, with the difference reaching its maximum near the pyramid
vertex, where $\Delta\varepsilon_{xx}\approx0.25\%$.

Islands with a non-uniform component content induce larger stretching strains
$\varepsilon_{xx}$ in the silicon substrate than uniform islands do. The
maximum difference is reached near the heterojunction and amounts to
$\Delta\varepsilon_{xx}\approx0.2\%$. At the substrate depth $z\sim h$, this
difference is practically equal to zero.

The effect of non-uniformity also manifests itself in a similar way in the
profiles of the strain tensor component $\varepsilon_{zz}$ (Fig.~6). In
particular, in the lower third of the quantum dot height ($z<h/3$), stretching
strains in uniform islands turn out larger than the corresponding values in
non-uniform quantum dots, with the maximum difference attaining $\Delta
\varepsilon_{zz}\approx0.13\%$ near the base. In the upper section of
nanoislands, the inverse relation takes place; namely, non-uniform islands
undergo larger stretching strains, and the maximum difference equals
$\Delta\varepsilon_{zz}\approx0.14\%$ near the vertex. In contrast to the case
of the tensor components $\varepsilon_{xx}$ and $\varepsilon_{yy}$, the difference
between the components $\varepsilon_{zz}$ for the considered types of quantum
dots survives at distances $z\sim4h$ in the substrate depth.

The obtained spatial dependences $\varepsilon_{ii}(x,y,z)$ $(i=$
$=x,y,z)$ were used
to evaluate the average strain values over the quantum dot volume $V_{\rm QD}$,%
\begin{equation}
\left\langle \varepsilon_{ii}\right\rangle =\int\limits_{V_{\rm
QD}}{\varepsilon _{ii}(x,y,z)dV}.
\end{equation}
The results of calculations showed that the nanoislands with a
uniform component content turn out more strained than the
non-uniform ones. The average strain values in them are
$\left\langle \varepsilon_{zz}\right\rangle \approx0.66\%,$ and
$\left\langle \varepsilon_{xx}\right\rangle \approx-1.4\%$ in
uniform islands against $\left\langle \varepsilon_{zz}\right\rangle
\approx$ $\approx0.63\%$ and $\left\langle
\varepsilon_{xx}\right\rangle \approx -1.1\%$ in non-uniform ones.
The origin of such differences consists in weaker local strains in
non-uniform quantum dots in the region near their bases, which
provides the main contribution at the averaging.

As was indicated above, strains can cause changes of the energy bands in
heterostructures. The results obtained in this work testify that the main
differences between the elastically deformed states of uniform and non-uniform
nanoislands are observed in their bulk. Therefore, the largest variations of
the energy structure are to be expected in quantum dots. In heterojunctions of
the second type, to which the junction Si--Ge belongs (see the inset in
Fig.~7), it is holes that are localized in the quantum dots. Therefore,
the non-uniformity in this region has to reveal a dominating influence on the
valence band bottom. To confirm this conclusion, the energy bands in the
examined heterostructure were calculated in the deformation potential
approximation. The change in the valence band bottom induced by
strains was calculated according to the relation%
\begin{equation}
\delta E_{V}=aS_{h}-\frac{bS_{b}}{2}+\frac{\Delta_{0}}{3},
\end{equation}
where $a$ and $b$ are the constants of the deformation potential for the valence
band, $\Delta_{0}$ is the spin-orbit splitting, and $S_{h}$ and $S_{b}$ are
the hydrostatic and biaxial strains, respectively, which are defined as
follows:
\begin{equation}
S_{h}=\varepsilon_{xx}+\varepsilon_{yy}+\varepsilon_{zz},
\end{equation}\vspace*{-7mm}%
\begin{equation}
S_{b}=2\varepsilon_{zz}-\varepsilon_{xx}-\varepsilon_{yy}.
\end{equation}

\begin{figure}
\vskip1mm
\includegraphics[width=7cm]{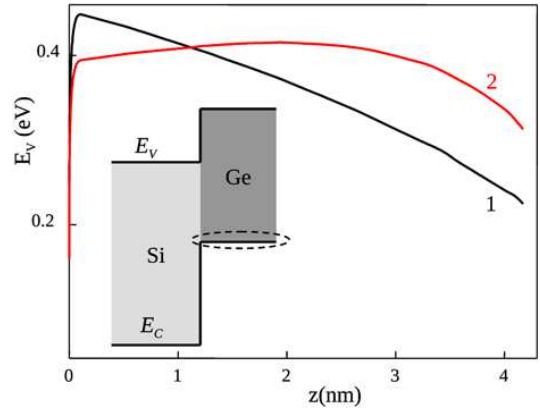}
\vskip-3mm\caption{Profiles $E_{V}(z)$ of the valence band bottom
energy in nanoislands with uniform (\textit{1}) and non-uniform
(\textit{2}) component contents. The inset illustrates the band
diagram of the Si$_{1-x}$Ge$_{x}$ heterojunction; the region
depicted in the main figure is outlined by a dashed curve  }
\end{figure}

The results obtained testify (Fig.~7) that, in nanoislands with a uniform
germanium content distribution, misfit strains give rise to an increase of
the valence band bottom near the heterojunction Si--Si$_{1-x}%
$Ge$_{x}$ (curve~\textit{1}). Therefore, the maximum of hole density
is observed in those quantum dots near the pyramidal island base. In
non-uniform nanoislands, the maximum of $E_{V}$ is attained
approximately at the middle of their height (curve~\textit{2}).
Hence, in quantum dots with a non-uniform content distribution, one
should expect that a redistribution of the hole concentration would
take place with a shift of its maximum toward the pyramid vertex.
Moreover, the changes in the dependence $E_{V}(x,y,z)$ induced by
the content non-uniformity would also stimulate modifications in the
energy spectrum of charge carriers, which will make a contribution
to the formation of the properties of
heterostructures.

\section{Conclusions}

To summarize, the elastic strain fields in SiGe heterostructures with quantum dots
synthesized following the Stranski--Krastanov growth mode were calculated. The
results of calculations revealed a difference between the elastically deformed
states of nanoislands with uniform and non-uniform distributions of silicon
and germanium contents over their volume. The calculations testify that the non-uniformity of the
contents of components changes the
spatial distributions of strain tensor components and diminishes their magnitudes in quantum dots. Using the
modification of the valence band bottom energy as an example, the influence of
the redistribution of germanium in nanoislands on the properties of SiGe
heterostructures is demonstrated.

\vskip3mm

{\it The work was sponsored by the State Fund for Fundamental
Researches (project F44, a \textquotedblleft Grant of the President
of Ukraine to support scientific researches of young scientists
in 2012\textquotedblright).}

\vspace*{3mm}
\rezume{%
В.В. Курилюк}{ПРУЖНІ ДЕФОРМАЦІЇ\\ В SiGe-ГЕТЕРОСТРУКТУРАХ З
КВАНТОВИМИ\\ ТОЧКАМИ НЕОДНОРІДНОГО СКЛАДУ} {Методом скінченних
елементів розраховано розподіли пружних деформацій в
гетероструктурах SiGe з квантовими точками. Досліджено вплив
неоднорідного розподілу германію всередині наноострівців на
просторові залежності та величину пружних полів. Показано, що
квантові точки сталого складу характеризуються більшими напруженнями
порівняно з неоднорідними наноострівцями.}

\end{document}